\documentclass[12pt,thmsa]{article}
\usepackage{amssymb}

\usepackage{sw20lart}

\input{tcilatex}
\begin{document}

\title{DARK\ MATTER\ OR\ A\ NEW\ FORCE?}
\author{R. L. Ingraham and L. A. Cummings \\
\bigskip \\
Department of Physics, MSC 3D\\
New Mexico State University \\
Las Cruces, NM 88003}
\date{}
\maketitle

\begin{abstract}
We show that a new force, which appeared in a five-dimensional
generalization of general relativity, implies precisely the flat circular
velocity curves of luminous matter observed in the outer parts of spiral
galaxies. This is attributed nowadays to a halo of dark matter exerting a
Newtonian force on the luminous matter. The same result could also be
obtained if general relativity were arbitrarily supplemented by a similar
4-tensor new force. Several other properties of this new force are
discussed. First, it acts also on light and enhances the usual gravitational
lensing by the luminous matter according to general relativity. Thus it acts
like the hypothetical dark matter in this respect also. Second, an
intriguing feature of the new force is that its energy tensor enters the
field equations for the metric with a sign opposite to that of ordinary
matter. Hence a cosmological solution for the new force (not the galactic
solution treated in this paper) might explain the cosmic negative, or
repulsive, gravitation responsible for the accelerated expansion of the
universe, attributed nowadays to ``dark energy'' and sometimes modelled by
Einstein's cosmological constant.

\bigskip

\textit{Subject headings:} galaxies --- dark matter --- flat rotation curves
--- new force.
\end{abstract}

\begin{center}
\pagebreak

\bigskip 1. INTRODUCTION
\end{center}

\bigskip

That the circular velocity $v_c$ of luminous matter departs from its
Keplerian form in the outer parts of many spiral galaxies and approximately
approaches a limit with distance (``flat rotation curve'') has been known
for about thirty years. This limiting value (Tully-Fisher relation) is 
\begin{equation}
v_c=220(L/L_{*})^{0.22}\text{ }km\text{ }s^{-1}\text{ ,}  \tag{1}
\end{equation}
where $L$ is the galaxy luminosity and $L_{*},$ a certain standard
luminosity. Nowadays this is generally ascribed to dark matter distributed
in a halo around the galaxy and exerting a Newtonian force on the luminous
matter. What this postulated dark matter is is not clear even today. Some
experts, however, are not completely convinced by the dark matter
hypothesis. Peebles for example (Peebles 1993) writes ``But since this
subject is still being explored, it is well to bear in mind the alternative
that we are not using the right physics.''

A new force, which turned up first in a 5-D (five-dimensional) version of
general relativity (Ingraham 1979), enforces precisely this flat circular
velocity curve. This calculation is given below. This new force $f_{\alpha
\beta }=-f_{\beta \alpha }$ is formally just like the electromagnetic force
tensor, but is an extra form of gravitation beyond that given by the metric.
Some details of this theory are given in \S 2.$^1$

This new force also acts on light, and produces a deflection of light
passing through or around a galaxy in addition to that due to the mass of
the galaxy predicted by GR (general relativity hereafter). So it would also
act like the dark matter halo in this respect. However, a new feature is the
following: this extra deflection does not have the full rotational symmetry
of the GR deflection. This can be seen intuitively because the new force
acts formally like the electromagnetic force in the motion equations of
matter and light. We plan to examine the data on gravitational lensing to
see if such an asymmetry shows up.

An intriguing feature of this new force is that it enters the field
equations for the metric with the ``wrong'' sign, that is, like a sort of
negative energy as opposed to the positive energy of ordinary matter or
radiation. This sign is unambiguous because the new force's
stress-energy-momentum tensor $T_{\alpha \beta }$ enters the metric field
equations without any coupling constant, to be contrasted with the
stress-energy-momentum of ordinary matter, which is coupled to the metric by
Newton's gravitational constant. \ This was noted in the original article as
anomalous (Ingraham 1979, bottom p. 243). However, today such a new force
should be quite welcome as a candidate for the so-called ``dark energy''
(Cowan 2001). Whether some cosmological solution for $f_{\alpha \beta }$
could explain the observed accelerating expansion of the universe is
difficult as yet to say because of the difficulty of solving the
cosmological field equations. These are now \textit{partial} differential
equations in the two independent variables $x^0\equiv $ time and $x^5\equiv
\lambda $, the fifth coordinate.$^2$ But the theory satisfies one stringent
necessary condition precisely because it is five-dimensional. Namely, due to
the isotropy demand for cosmological solutions, no purely space-time
components $f_{\mu \nu }$, $\mu $, $\nu =0,1,2,3,$ could survive because
they obviously would define preferred directions in the spatial universe.
But in five dimensions we can look for a solution with only $f_{05}\neq 0$.

\bigskip

\begin{center}
2. DERIVATION\ OF\ THE\ FLAT CIRCULAR\ VELOCITY\ CURVE
\end{center}

\bigskip

We model the spiral galaxy as a spherically symmetric core of mass M with
negligible mass outside the core. The field equations for the 5-D metric $%
\gamma _{\alpha \beta }$ outside the core (Ingraham 1979, eqs. (2.19a,b) or
(2.21a,b)) will not be needed here. The field equation for the new force $%
f_{\alpha \beta }$ outside the core is $\bigtriangledown _{_{_{_\alpha
}}}f^{^\alpha }$ $_{_{_{_\beta }}}=0$, where $\bigtriangledown _{_{_{_\alpha
}}}$ is the covariant derivative, or equivalently 
\begin{equation}
\partial _{_{_{_\alpha }}}(\sqrt{\gamma }\text{ }f^{\alpha \beta })=0\text{
, \qquad }\gamma \equiv \text{ det }\gamma _{_{\alpha \beta }}\text{ .} 
\tag{2}
\end{equation}
For this galactic problem we assume that $f_{\alpha \beta }$ is small of
first order. Then it perturbs the metric only to second order, since $%
T_{\alpha \beta }$ is quadratic in $f_{\alpha \beta }$. Hence we can use a
spherically symmetric O-order line element of Schwarzschild form: 
\begin{equation}
d\Theta ^2=-\lambda ^{-2}g_{\alpha \beta }\text{ }dx^{^\alpha }dx^{^\beta }%
\text{ ,\qquad }\alpha ,\beta =0,1,2,3,5\text{ ,\qquad }  \tag{3a}
\end{equation}
\[
g_{\alpha \beta }\text{ }dx^{^\alpha }dx^{^\beta }=e^{2\mu }\text{ }%
dr^2+r^2(d\theta ^2+\sin ^2\theta \text{ }d\phi ^2)-e^{2\nu }dt^2+e^{2\xi }%
\text{ }d\lambda ^2\text{ ,}
\]
\begin{equation}
x^1\equiv r\text{ , }x^2\equiv \theta \text{, }x^3=\phi \text{, }%
x^0=t,x^5\equiv \lambda \quad (\text{units: }c=1\text{) ,}  \tag{3b}
\end{equation}
where the metric coefficients $\mu $, $\nu $, and $\xi $ are functions of $r$
and $\lambda $ only.

Here $d\Theta $ has the geometrical meaning of the infinitesimal angle of
intersection of the space-time spheres $x^{^\alpha }$ and $x^{^\alpha
}+dx^{^\alpha }$ where $x^{^\mu }$, $\mu =0,1,2,3$, is the center and $%
x^5\equiv \lambda $ is the radius of this sphere (Ingraham 1998). The
15-parameter conformal group is the symmetry group of this angle metric in
the gravitation-free case. It is convenient to factor a $-\lambda ^{-2}$ out
of the angle metric and deal with the $g_{\alpha \beta }$ in most problems
because the $g_{\alpha \beta }$ reduce to constants in Cartesian coordinates
in the gravitation-free world.

The ordinary Schwarzschild metric in GR, 
\begin{equation}
e^{2\nu }=e^{-2\mu }=1-2m/r\text{ , }e^{2\xi }=1\text{ ,}  \tag{4}
\end{equation}
is an exact solution of the metric field equations with $f_{\alpha \beta
}\equiv 0$. Here $m\equiv GM\diagup c^2$, $M\equiv $ mass of the galactic
core, is the geometric mass. It makes sense to use this $\lambda $%
-independent solution for the $g_{\alpha \beta }$ here because a $\lambda $%
-dependence of the $g_{\alpha \beta }$ is forced only by $T_{\alpha \beta }$%
, and we are neglecting that here.

The motion equations are in general 
\begin{equation}
\frac{d^2x^{^\alpha }}{d\Theta ^2}+\left\{ 
\begin{array}{l}
\;\alpha  \\ 
\beta \gamma 
\end{array}
\right\} \frac{dx^{^\beta }}{d\Theta }\text{ }\frac{dx^{^\nu }}{d\Theta }%
-f^{^{^\alpha }}\text{ }_{_\beta }\text{ }\frac{dx^{^\beta }}{d\Theta }=0%
\text{ .}  \tag{5}
\end{equation}
Note the universal form of eq. (5): there is no dimensional strength
constant multiplying $f^{^\alpha }$ $_{_\beta }$ $dx^{^\beta }\diagup
d\Theta $, so that \textit{in the angle geometry} the new force accelerates
all bodies the same, just as does the force from the metric. This form is
mathematically forced. Since $T_{\alpha \beta }$ occurs in the metric field
equations without a coupling constant, $f_{\alpha \beta }$ must have
dimensions (length)$^{-2}$ in cartesian coordinates (we write this $%
f_{\alpha \beta }\thicksim $ (length)$^{-2}$ hereafter). Thus \newline
$f^{^\alpha }$ $_{_\beta }\equiv \gamma ^{\alpha \delta }f_{\delta \beta }$
is dimensionless. This fact plus the fact that $d\Theta $ is dimensionless
forbids any dimensional strength constant, as asserted.

To compare with present day motion equations it is convenient (in fact
necessary) to introduce a length variable $\tau $ via $d\tau /d\Theta
=\lambda ^2\diagup \ell $, where $\ell $ is a length parameter (Ingraham
1998). This $\tau $ turns out to be the proper time in simple cases,
including the present case (see later). The motion equations (5) with the
Schwarzschild metric (3) can be rewritten with $\tau $ the independent
variable. We note that the driving term in the new force then appears as 
\newline
$\ell f^{\underline{\alpha }}$ $_{_\beta }$ $dx^{^\beta }\diagup d\tau $ in
the left members, where $f^{\underline{\alpha }}$ $_{_\beta }\equiv
g^{\alpha \gamma }f_{\gamma \beta }$. Of these we need only the two angle
motion equations here, 
\begin{equation}
\begin{array}{ll}
\stackrel{\bullet \bullet }{\phi }+2\stackrel{\bullet }{r}\stackrel{\bullet 
}{\phi }/r\text{ }+2\cot \theta \stackrel{\bullet }{\phi }\stackrel{\bullet 
}{\theta }+\ell \text{ }f^{\underline{3}}\text{ }_{_\beta }\text{ }\stackrel{%
\bullet }{x}^{^\beta } & =0\text{ ,}
\end{array}
\tag{6a}
\end{equation}
\begin{equation}
\begin{array}{ll}
\stackrel{\bullet \bullet }{\theta }+2\stackrel{\bullet }{r}\stackrel{%
\bullet }{\theta }\stackrel{.}{\text{ }/r-\sin \theta \cos }\theta \stackrel{%
\bullet }{\phi }^2+\ell \text{ }f^{\underline{2}}\text{ }_{_\beta }\text{ }%
\stackrel{\bullet }{x}^{^\beta } & =0\text{ ,}
\end{array}
\tag{6b}
\end{equation}
where the dot means $d/d\tau $ . Choose the polar coordinates such that the
galactic core's center is at the origin and the disc of the galaxy lies in
the plane $\theta =\pi /2$. Then we hypothesize that the new force has only
the component $f_{13}(r,\theta ,\lambda )\neq 0$. The field equations (2)
for the Schwarzschild metric (3) with $\sqrt{\gamma }=\lambda ^{-5}e^{\mu
+\nu +\xi }$ $r^2\sin \theta $ and $f^{\alpha \beta }=\lambda ^4$ $f^{%
\underline{\alpha }\underline{\beta }}$ reduce to that for $\beta =3$, which
is 
\begin{equation}
\partial _1(\lambda ^{-1}r^2\sin \theta \text{ }f^{\underline{1}\text{%
\underline{$3$}}}\text{ })=0\text{ .}  \tag{7}
\end{equation}
The solution is 
\begin{equation}
f^{\underline{1}\text{\underline{$3$}}}=C(\lambda ,\theta )\diagup r^2\text{
,\qquad }f_{13}=C(\lambda ,\theta )\text{ }e^{2\mu }\sin ^2\theta \text{ ,} 
\tag{8}
\end{equation}
where $C(\lambda ,\theta )$ is arbitrary. We choose $C(\lambda ,\theta
)=C\diagup \sin ^2\theta $ , $C=const$. Then the $\phi $-motion (6a) has the
driving term $-\ell Ce^{2\mu }\stackrel{.}{r}\diagup r^2\sin ^2\theta $.
Multiply eq. (6a) by $r^2\sin ^2\theta $ and integrate once. The result can
be written in the form

\begin{equation}
\sin ^2\theta \;\;r\stackrel{\bullet }{\phi }=\ell CG(r)+h/r\text{ ,} 
\tag{9a}
\end{equation}
\begin{equation}
G(r)\equiv 1-2m/r+(2m/r)\log (r/2m-1)\text{ , }h=const.\text{ }\qquad . 
\tag{9b}
\end{equation}
For the $\theta $-motion (6b) $f^{\underline{2}}$ $_{_\beta }\equiv 0$, but
there is no integrating factor. However, $\theta =\pi /2$ is a solution. The 
$t$-motion integrates to $t=ae^{-2\nu }$, $a=$ const. The $\lambda $-motion
can be integrated fully, with the result $\lambda =(\tau ^2+\ell ^2)^{1/2}$,
and implies 
\begin{equation}
\stackrel{\bullet }{\lambda }^2+\ell ^2/\lambda ^2=1\text{ .}  \tag{10}
\end{equation}
If eq. (3) is rewritten to give $d\tau ^2$ and the result (10) is used, it
is seen that $\tau $ is precisely the proper time. As to the $r$-motion, see
below.

So for the motion of material in the galactic disc $\theta =\pi /2$ we have
from eq. (9) that the circular velocity $v_c$ is 
\begin{equation}
v_c\equiv rd\phi /dt=r\stackrel{\bullet }{\phi }/\stackrel{\bullet }{t}\text{
}=\ell C/a+h/ar\text{ }(\text{units: }c=1)  \tag{11}
\end{equation}
approximately, where we have put $G(r)\approx 1$ and $e^{2\nu }\approx 1$
since $C$ is already assumed first order small and $2m/r<<1$.$^3$ Thus $%
v_c\approx \ell C/a=$ const. if $\ell C$ dominates the usual angular
momentum term $h/r$ at the distances of interest, and certainly goes to a
constant as $r\rightarrow \infty $. Taking $v_c=220$ km $s^{-1}$ for
example, one gets $\ell C\approx v_c/c\approx 7.3\times 10^{-4}$ for the
dimensionless combination $\ell C$ for nonrelativistic material motion $%
(a\approx 1)$.

About the $r$-motion we mention here only that the orbit will differ from
the GR result of an ellipse with precession of perihelion (Adler et al.1975)
if the new force term $\ell C$ in eq. (11) dominates $h/r$, as observations
suggest.

\bigskip

\begin{center}
3. CONCLUDING\ REMARKS
\end{center}

\bigskip

\noindent 1.\qquad This explanation of the flat circular velocity curve of
luminous matter in the galactic disc of spiral galaxies could be phrased in
terms of GR equipped with an extra new force of gravitation $F_{\mu \nu
}(\mu ,\nu =0,1,2,3)$ without recourse to the 5-D theory as follows.

Take the usual Schwarzschild solution for the $O$-order 4-D metric $g_{\mu
\nu }$ outside the core. The field equations for $F_{\mu \nu }$ would be 
\begin{equation}
\partial _{_\mu }(\sqrt{g}\text{ }F^{\mu \nu })=0\text{ , }\sqrt{g}=r^2\sin
\theta   \tag{12}
\end{equation}
in analogy to eq. (2). Assuming only $F_{13}(r,\theta )\neq 0$ , we get the
solution 
\begin{equation}
F^{13}=A(\theta )\diagup r^2\text{ , }F_{13}=A(\theta )\text{ }e^{2\mu }\sin
^2\theta \text{ ,}  \tag{13}
\end{equation}
where $A(\theta )$ is arbitrary. Choose $A(\theta )=A\diagup \sin ^2\theta $%
, $A=const.$ The $\phi $-motion equation is identical to eq. (6a) except
that the driving term is $\ell F^3$ $_{_\nu }$ $\stackrel{.}{x}^{^\nu }$,
and the dot means $d\diagup d\tau $, $\tau \equiv $ proper time.$^4$
Inserting the solution (13), one integrates once to get 
\begin{equation}
\sin ^2\theta \;\;r\stackrel{\bullet }{\phi }=\ell AG(r)+h/r\text{ ,} 
\tag{14}
\end{equation}
in strict analogy with eq.(9).\ The $\theta $-motion again yields the
particular solution $\theta =\pi /2$, valid for motion in the galactic disc.
The $t$-motion is as before. Hence we get the result 
\begin{equation}
v_c\approx \ell A/a+h/ar\qquad (units\text{: }c=1)\text{ ,}  \tag{15}
\end{equation}
or a flat circular velocity curve when $\ell A$ dominates $h/r$.

While this theory may be more attractive to some theorists because it stays
within the bounds of four dimensions and uses the familiar, accepted theory
of general relativity, it is open to the charge of being \textit{ad hoc}:
there is no \textit{mathematical} necessity of grafting a new field $F_{\mu
\nu }$ onto the beautifully economical structure of GR. The interested
reader may well wonder whether the same objection applies to the 5-D theory
presented here. The answer is ``No''. If we had started with a 5-D \textit{%
Riemannian} space of metric $\gamma _{\alpha \beta }$, then equipping it
with the extra field $f_{\alpha \beta }$ would have been equally \textit{ad
hoc}. But the field equations of this theory (Ingraham 1979) were derived
from varying a simple action of Einsteinian form in a 5-D \textit{projective}
space furnished with a $6\times 6$ quadric $S_{\rho \tau }$, a function of
six ``homogeneous'' coordinates $X^{^\rho }(\rho ,\tau =1,2,\cdots 6)$. The
resulting field equations for $S_{\rho \tau }$ were then transformed into
equations with respect to the five ``inhomogeneous'' coordinates $x^{^\alpha
}$, and the split into the 5-D Riemannian tensors $\gamma _{\alpha \beta }$
and $f_{\alpha \beta }$ occurred as a result of this transformation. This
derivation and transformation are given in detail in the cited work
(Ingraham 1979, pp. 236-241).

\noindent 2.\qquad The motion equation (5) is clearly universal: all bodies
are accelerated the same, there is no constant referring to the particular
body in question (its mass, charge, etc.) multiplying $f^{^\alpha }$ $%
_{_\beta }$ $dx^{^\beta }\diagup d\Theta $ . As we showed, dimensions forbid
it. However, when we go to the motion in terms of the proper time $\tau $
via the substitution $d\tau /d\Theta =\lambda ^2/\ell $, a dimensional
strength constant $\ell $, a length, appears in the motion, \textit{cf}.
eqs. (6).\ This $\ell $ may refer to the properties of the particular body
in question or it may be a universal length; which alternative holds is not
clear at the moment. However, the result $v_c/c\approx \ell C$ $(a\approx 1)$
shows that this $\ell $ is the same at least for all the forms of luminous
matter which obey the Tully-Fisher relation\ (1).

\bigskip

\begin{center}
REFERENCES
\end{center}

\bigskip

\noindent Adler, R., Bazin, M. \& Schiffer, M. 1975, Introduction to General
Relativity (New York: McGraw Hill)

\medskip

\noindent Cowen, R. 2001, Science News, 159, 196

\medskip

\noindent Ingraham, R.L. 1979, Il Nuovo Cimento, 50B, 233 

\noindent \_\_\_\_\_\_\_\_\_1988, Int. J. Mod. Phys., 7, 603

\medskip

\noindent Peebles, P.J.E. 1993, Principles of Physical Cosmology (Princeton,
N.J.: Princeton University Press), 47

\bigskip

\bigskip

\begin{center}
FOOTNOTES
\end{center}

\bigskip

\noindent $^1\qquad $This 5-D theory is enforced if conformal group
symmetry, which generalizes the present day Poincar\'{e} group symmetry, is
demanded as the fundamental kinematical symmetry of physics (Ingraham 1998).
It has nothing to do with Kaluza or Kaluza-Klein theories, has very
different physical consequences, etc.

\bigskip

\noindent $^2$\qquad This fifth coordinate has a well-defined role in the
dynamics of particles (Ingraham 1998) and should be observable, unlike the
hidden extra dimensions in some current particle theories.

\bigskip

\noindent $^3$\qquad If we take $M=3-5\times 10^{10}$ $M_{\odot }$ for the
mass of the core and $r>5$ kpc for a typical flat circular velocity curve
(Peebles 1993, cf. Fig. 3.12), one finds $2m/r<2-3\times 10^{-7}$.

\bigskip

\noindent $^4$\qquad The length strength constant $\ell $ must appear here
by dimensions if we take $F_{\mu \nu }\thicksim $ $(length)^{-2}$ like $%
f_{\alpha \beta }$.

\end{document}